\documentclass[review]{elsarticle}

\usepackage{lineno,hyperref}

\journal{arXiv}

\usepackage{amssymb,bm,amsbsy,graphicx,color,amsmath,amsfonts}

\usepackage{geometry}
\newcommand{\posCur}{\vect{x}}
\newcommand{\posIni}{\vect{X}}
\newcommand{\posmidCur}{\vect{\varphi}}
\newcommand{\posmidIni}{\vect{\varphi}^{(0)}}
\newcommand{\ConvMembCur}{\bf{b}}

\newcommand{\ConvMembIni}{{\bf{B}}}

\newcommand{\gMembCur}{{\bf{a}}}
\newcommand{\gMembCurComp}{{a}}
\newcommand{\gShellCur}{{\bf{g}}}
\newcommand{\gShellCurcomp}{{g}}

\newcommand{\gmidShellCur}{{\bf{a}}}

\newcommand{\gMembIni}{{\bf{A}}}
\newcommand{\gShellIni}{{\bf{G}}}
\newcommand{\gShellInicomp}{{G}}

\newcommand{\gmidShellIni}{{\bf{A}}}
\newcommand{\gmidShellCurcomp}{{a}}
\newcommand{\gmidShellInicomp}{{A}}
\newcommand{\CMemb}{\bar{\bf{C}}}
\newcommand{\CShell}{{\bf{C}}}

\newcommand{\coord}{\xi}
\newcommand{\coordOrth}{s}
\newcommand{\FMemb}{\bar{\bf{F}}}
\newcommand{\FShell}{{\bf{F}}}

\newcommand{\thickIni}{H}

\newcommand{\curvCurcomp}{{\kappa}}
\newcommand{\curvInicomp}{{\kappa}^{(0)}}

\newcommand{\changeCurvcomp}{{\chi}}
\newcommand{\JShell}{{J}}

\newcommand{\vect}[1]{\boldsymbol{#1}}

\newcommand{\invivo}{\emph{in vivo }}

\newcommand{\CauchyStress}{\mathbf{\sigma}}
\newcommand{\stiffsupp}{k_s}
\newcommand{\dampsupp}{k_d}
\newcommand{\thoracicpressure}{p_0}
\newcommand{\disp}{\mathbf{u}}

\newcommand{\velocity}{\mathbf{v}}

\newcommand{\firstmat}{c_1}
\newcommand{\secondmat}{c_2}
\newcommand{\isomat}{c}
\newcommand{\firstInv}{I_C}
\newcommand{\secondInv}{II_C}
\newcommand{\fiberstretch}{\lambda}
\newcommand{\fiberdirection}{\mathbf{M}}
\newcommand{\fiberangle}{\alpha_0}

\newcommand{\esef}{\psi_\textrm{el}}
\newcommand{\tsef}{\psi}
\newcommand{\lmShell}{p}

\newcommand{\etal}{\emph{et al.}}

\begin{document}

\begin{frontmatter}

\title{A Systematic Comparison between Membrane, Shell, and 3D Solid Formulations for Non-linear Vascular Biomechanics}

\author{Nitesh Nama$^{a,1}$, Miquel Aguirre$^{b,c,d}$, Rogelio Ortigosa$^e$, Antonio J. Gil$^f$, Jay D. Humphrey$^g$, \\C. Alberto Figueroa$^{h,i}$}
\fntext[]{Email: nitesh.nama@unl.edu}


\address[mymainaddress]{Department of Mechanical \& Materials Engineering, University of Nebraska-Lincoln, NE, USA}
\address[mymainaddress]{Laboratori de Càlcul Numèric, Universitat Politècnica de Catalunya, Jordi Girona 1, E-08034, Barcelona, Spain}
\address[mymainaddress]{International Centre for Numerical Methods in Engineering (CIMNE), Gran Capità, 08034, Barcelona, Spain}
\address[mymainaddress]{Mines Saint-Étienne, Univ Lyon, Univ Jean Monnet, INSERM, U 1059 Sainbiose, Centre CIS, F - 42023 Saint-Étienne, France}
\address[mysecondaryaddress]{Computational Mechanics and Scientific Computing Group, Technical University of Cartagena, \\Campus Muralla del Mar, 30202, Cartagena (Murcia), Spain}
\address[mymainaddress]{Zienkiewicz Centre for Computational Engineering, College of Engineering, Swansea University, \\Bay Campus, SA2 0SN, UK}
\address[mymainaddress]{Department of Biomedical Engineering, Yale University, New Haven, CT, USA}
\address[mymainaddress]{Department of Surgery, University of Michigan, Ann Arbor, MI, USA}
\address[mymainaddress]{Department of Biomedical Engineering, University of Michigan, Ann Arbor, MI, USA}
\begin{abstract}
Typical computational techniques for vascular biomechanics represent the blood vessel wall
via either a membrane, a shell, or a 3D solid element. Each of these formulations has its trade offs concerning accuracy, ease of implementation, and computational costs.
Despite the widespread use of these formulations, a systematic comparison on the performance and accuracy of these formulations for nonlinear vascular biomechanics is lacking. Therefore, the decision regarding the optimal choice often relies on intuition or previous experience, with unclear consequences of choosing one approach over the other. Here, we present a systematic comparison among three different formulations to represent vessel wall: (i) a nonlinear membrane model, (ii) a nonlinear, rotation-free shell model, and (iii) a nonlinear 3D solid model. For the 3D solid model, we consider two different implementations employing linear and quadratic interpolation. We present comparison results in both idealized and subject-specific mouse geometries. For the idealized cylindrical geometry, we compare our results against the axisymmetric solution for three different wall thickness to radius ratio. Subsequently, a comparison of these approaches is presented in an idealized bifurcation model for regionally varying wall thickness. Lastly, 
we present the comparison results for a subject-specific mouse geometry with regionally varying material properties and wall thickness, while incorporating the effect of external tissue surrounding the vessel wall.

\end{abstract}


\end{frontmatter}


\section{Introduction}
Computational techniques to simulate cardiovascular biomechanics in three-dimensional (3D) models of arteries have attracted significant interest owing to their applications in disease research, medical device design, and surgical planning~\cite{lally2005cardiovascular,figueroa2009computational,laubrie2020new,raghavan2000wall,humphrey2013cardiovascular,maas2012febio}. Computational models have utilized various representation of the vessel wall which has been modeled, with varying degree of detail, as either a 3D solid or a two-dimensional (2D) structure (membrane or shell). The latter choice has been motivated by the fact that most arteries and veins are characterized by a low thickness to radius ratio, rendering them suitable to a 2D representation. Nonetheless, each of these approaches has its own advantages and limitations concerning computational efficiency, cost, and ease of implementation. 

In general, 3D solid models of vessel wall offer excellent accuracy, straightforward implementation, and flexibility to easily incorporate arbitrary constitutive material models~\cite{wriggers1996comparison}. 3D solid models have been extensively employed both for the solid mechanics analysis of arterial wall~\cite{doyle2007comparison, weisbecker2014generalized, liu2017new, lally2005cardiovascular,riveros2013pull} as well as for fluid-structure analyses~\cite{gerbeau2005fluid,colciago2014comparisons,bazilevs2010fully}. However, 3D solid models employing low-order interpolation present significant challenges concerning various locking phenomena and difficulty in generating good quality meshes for thin geometries. Specifically, generating sufficiently good mesh (i.e. with appropriate aspect ratio of mesh elements) for thin geometries requires a finer in-plane mesh, leading to significant increased degrees of freedom. This challenge can, in principle, be circumvented via the use of higher-order interpolation in 3D solid formulations. However, such formulations are typically undesirable for vascular applications owing to their associated instabilities and high computational cost.

To circumvent these challenges, 2D manifold representations of the vessel wall have been proposed. Earlier works have employed a membrane representation of the vessel wall and have highlighted its significantly low computational cost. Here, the vessel wall has been modeled as either a linear~\cite{figueroa2006coupled} or a nonlinear membrane~\cite{lu2008inverse,kyriacou1996finite}. More recently, various shell models have been employed to incorporate the bending behavior of the vessel wall~\cite{kim2008dynamic,tepole2015isogeometric, zhou2010patient, laubrie2020new, martin2015patient,Nama2020}. 
While shell theory is a mature field in nonlinear mechanics
~\cite{bischoff2004models}, the adaptation of these theories for vascular biomechanics is largely restricted to shell formulation that employ Kirchhoff-Love theory. In particular, we note the recent stress resultant shell formulation employed by Kim \etal~\cite{kim2008dynamic}(in the context of heart valve leaflets), the isogeometric formulation by Tepole \etal~\cite{tepole2015isogeometric}, and our recent work concerning rotation- free shell formulation~\cite{Nama2020}. 
We also note that subject-specific vascular geometric models are typically obtained from medical images, where a direct assessment of vessel wall thickness is difficult. Consequently, the construction of a 3D volumetric geometric model of vessel wall entails the creation of a surface mesh that is then extruded to obtain a 3D solid mesh. In this context, the 2D manifold representations of vessel wall offer another advantage over 3D models by avoiding the need for this additional pre-processing step.

However, despite the widespread use of 2D manifold models of vascular walls, there is a limited understanding of the comparison of their performance against 3D solid formulations~\cite{wriggers1996comparison,chen2004comparison}. To the best of authors' knowledge, there has been no comprehensive comparison of these approaches in the context of vascular biomechanics for complex subject-specific geometries and biologically relevant material models. Consequently, the modeling choice often relies on intuition and preference, with unclear consequences of choosing one approach over the other.

Another issue that introduces differences between 2D manifold models and 3D solid models concerns the handling of incompressibility condition for the vessel wall material. Typically, 3D solid models of vessel wall do not enforce exact incompressibility condition in order to avoid the introduction of a Lagrange multiplier. Instead, they choose to enforce \emph{near incompressibility} for the vessel wall materials. In contrast, 2D manifold representations of vessel wall enforce exact incompressibility for the vessel wall by expressing through-thickness quantities in terms of in-plane quantities. Again, the consequences of this difference in handling of incompressibility condition are unclear.

Here, we address this gap in the literature by presenting a systematic comparison between three different models of vessel wall. The first model is a nonlinear membrane model used previously by Lu~\etal~\cite{lu2008inverse} for cerebral aneurysms. The second is a nonlinear, rotation-free shell model, originally proposed by O{\~n}ate and Flores~\cite{onate2005advances}, and recently adapted for biomechanics by Nama~\etal~\cite{Nama2020}. The third model is a nonlinear 3D solid model, where we consider two different implementations employing wedge (or triangular prismatic) elements with linear and quadratic interpolation~\cite{zienkiewicz2005finite,yamakawa2009converting,afazov2012mathematical,meftah2021six}. 

The outline of rest of the article is as follows: In Section 2, we briefly describe each formulation with emphasis on assumptions concerning through-thickness variation of kinematic quantities. We also highlight the differences in the handling of incompressiblity condition among these approaches. We then describe our extrusion strategy to obtain 3D solid wedge elements from the 2D midsurface triangular elements to ensure a consistent comparison among different formulations. In Section 3, we describe the results of comparison among different formulations for three different geometrical models. We first consider an idealized cylindrical geometry to compare the results of each formulation against the axisymmetric theory and investigate the performance of different formulations as a function of wall thickness to radius ratio. We also compare the performance of different formulations in an idealized bifurcation model with regionally varying wall thickness. We then present the comparison results for a subject-specific mouse geometry with regionally varying material properties and wall thickness, while incorporating the effect of external tissue surrounding the vessel wall. Lastly, we provide a brief discussion of our results in Section 4.

\section{Theoretical Formulation}
In this section, we describe details of the three formulations considered in this article: \emph{(i)} a nonlinear membrane, \emph{(ii)} a nonlinear, rotation-free shell, and \emph{(iii)} a nonlinear 3D solid. Specifically, for each formulation, we consider the mapping associated with the deformation of a body from its material configuration to its deformed configuration. Subsequently, we utilize this mapping to provide expressions for the deformation tensor and the right Cauchy-Green strain tensor. These expressions are then employed to define the elastic strain energy for the two constitutive models considered in this work.

In the rest of the article, Greek subscripts or superscripts take values $\{1,2\}$ while Latin subscripts or superscripts take values $\{1,2,3\}$. We follow Einstein's index notation where a repeated symbol denotes a summation, unless stated otherwise. $\posCur$ and $\posIni$ denote a point in the continuum in current and initial configuration, respectively, while $\posmidCur$ and $\posmidIni$ denote a point on the midsurface of the continuum in current and initial configuration, respectively.

\subsection{Kinematics}

In this section, we describe the main notions regarding the kinematic description of the three formulations employed in this paper. For each formulation, our objective is to obtain expressions for the right Cauchy-Green strain tensor that can further be employed to define the elastic strain energy.

\subsubsection{Membrane Formulation}
In this section, we describe the nonlinear membrane formulation, previously employed by Lu~\etal~\cite{lu2007inverse} for cerebral aneurysms. A membrane formulations represents the simplest modeling choice for the vessel wall and allows the description of entire kinematics via mid-surface quantities~\cite{gruttmann1992theory}. 

\begin{figure}[!htbp]
	\centering
	\includegraphics[width=0.8\textwidth]{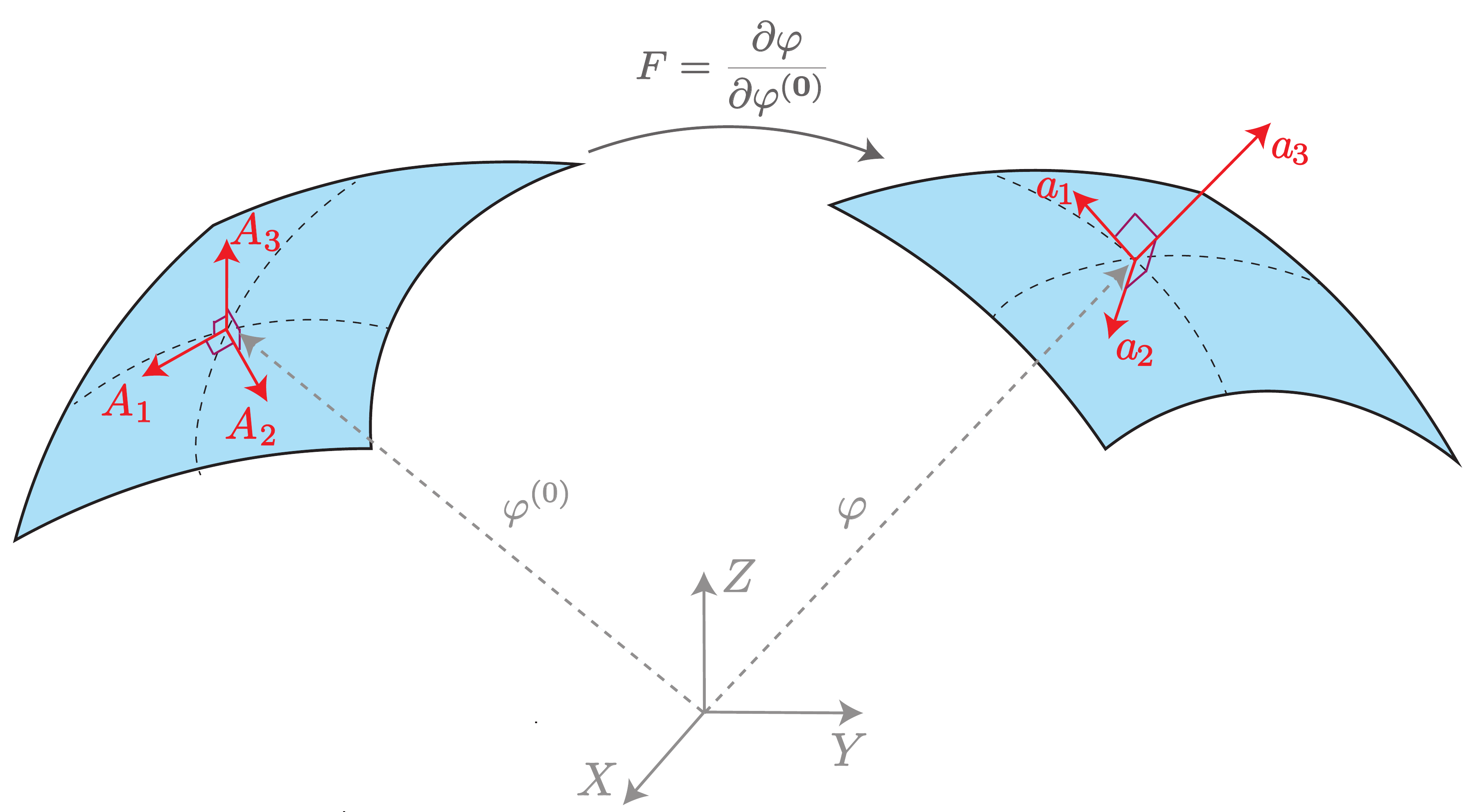}
	\caption{Schematic drawing illustrating the basis vectors in the initial and deformed configuration for the membrane formulation.}
	\label{fig:memb_schematic}
\end{figure}

Figure~\ref{fig:memb_schematic}
shows the material and the deformed configurations of the vessel wall midsurface. Consider a mapping between a material point $\posmidIni$ and a spatial point $\posmidCur$, with both points belonging to the membrane surface. 
The surface is parametrized by  convective coordinates $\coord^{\alpha}$ ($\alpha=1,2$), which yield a basis of two covariant vectors in the tangent planes of the initial and deformed configurations
\begin{equation}
\label{eq:membbasis}
\ConvMembIni_\alpha=\frac{\partial\posmidIni}{\partial\coord^\alpha},
\quad
\ConvMembCur_{\alpha}=\frac{\partial \posmidCur}{\partial\coord^{\alpha}}.
\end{equation}
\noindent Following Gruttman and Taylor~\cite{gruttmann1992theory}, we introduce an orthonormal basis in the material configuration; in this case the covariant and contravariant components of stress and strain tensors are identical, thereby simplfying the implementation. To this end, we define
%
\begin{equation}
\label{eq:orthonormal_basis}
\gMembIni_3=\frac{\ConvMembIni_1\times\ConvMembIni_2}{\|\ConvMembIni_1\times\ConvMembIni_2\|},
\quad
\gMembIni_1=\frac{\ConvMembIni_1}{\|\ConvMembIni_1\|},\quad\gMembIni_2=\gMembIni_3\times\gMembIni_1,
\end{equation}
\noindent with associated coordinates $s_\alpha$ indirectly defined as
\begin{equation}
\gmidShellIni_\alpha=\frac{\partial\posmidIni}{\partial \coordOrth_\alpha}.
\end{equation}
We remark that this orthonormal basis is introduced only for mathematical convenience and ease of implementation and is not strictly necessary to describe the membrane formulation. Similarly, the basis vectors (not necessarily orthonormal) at a position $\posmidCur$ of a point on the membrane surface in the current configuration can be defined as
\begin{equation}
\gmidShellCur_\alpha = \frac{\partial\posmidCur}{\partial\coordOrth_\alpha}. \label{eq: currentbases}
\end{equation}

\noindent The in-plane deformation gradient can be defined as
\begin{align}
\FMemb=\frac{\partial\posmidCur}{\partial\posmidIni}
=\frac{\partial\posmidCur}{\partial\coordOrth_\alpha}\otimes\frac{\partial\coordOrth_\alpha}{\partial\posmidIni}
=\gMembCur_\alpha\otimes\gMembIni^\alpha.\label{eq:def_gradient_membrane}
\end{align}
Correspondingly, the in-plane right Cauchy-Green strain tensor is obtained as
\begin{align}
\label{eq: CMemb}
\CMemb=\gMembCurComp_{\alpha \beta} \gMembIni^\alpha \otimes \gMembIni^\beta,
\end{align}
where $\gMembCurComp_{\alpha \beta}$ denotes the metric in the current configuration
\begin{align}
\label{eq: membcovariantmetric}
\gMembCurComp_{\alpha \beta}=\gMembCur_\alpha \cdot \gMembCur_\beta.
\end{align}
Equation~(\ref{eq: CMemb}) will be used to define the elastic strain energy for the membrane formulation in Section~\ref{sec:ConsModel}. We refer the reader to Lu~\etal~\cite{lu2007inverse} for further details on membrane formulation.

\subsubsection{Rotation-free Shell Formulation}
Next, we describe the rotation-free shell formulation, originally proposed by O\~{n}ate and Flores~\cite{onate2005advances} and recently adapted for biomechanics by Nama~\etal~\cite{Nama2020}. This shell formulation employs the Kirchhoff hypothesis of straight and normal cross-sections and describes the kinematics of vessel wall via its midsurface and associated normal vector. Below, we provide a brief overview of this formulation and refer the reader to our earlier work~\cite{Nama2020} for a more detailed description.

\begin{figure}[!htbp]
	\centering
	\includegraphics[width=0.8\textwidth]{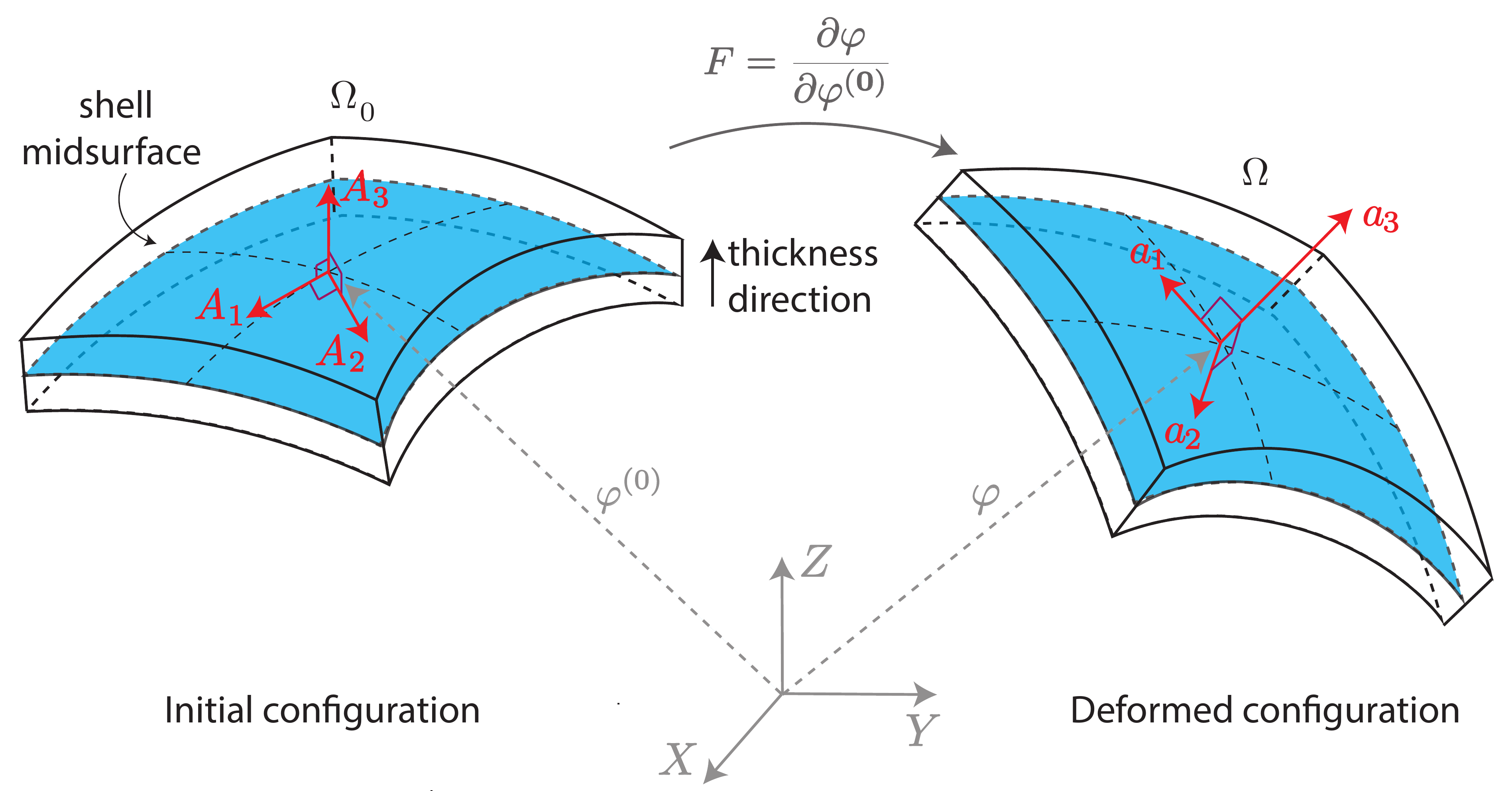}
	\caption{Schematic drawing illustrating the basis vectors in the initial and deformed configuration for the shell formulation.}
	\label{fig:shell_schematic}
\end{figure}

Consider a mapping as shown in Figure~\ref{fig:shell_schematic} where a material point $\posmidIni$ is mapped to a spatial point $\posmidCur$, with both points belonging to the shell midsurface. Similar to the membrane formulation described above, and following Gruttman and Taylor~\cite{gruttmann1992theory}, we define an orthonormal basis corresponding to a point on the shell midsurface in the initial configuration as
\begin{align}
\label{eq: shell_ortho}
    \gmidShellIni_\alpha=\frac{\partial\posmidIni}{\partial \coordOrth_\alpha},\quad\alpha=1,2.
\end{align}
with the unit normal vector to the midsurface
\begin{align}
    \gmidShellIni_3=\frac{\gmidShellIni_1\times\gmidShellIni_2}{\|\gmidShellIni_1\times\gmidShellIni_2\|}.
\end{align}
Again, as noted earlier in the discussion of membrane formulation, this orthonormal basis is introduced only for mathematical convenience and ease of implementation and is not strictly necessary to describe the shell formulation.
Similar to the initial configuration, the basis vectors (not necessarily orthonormal) at a position $\posmidCur$ of a point on the shell midsurface in the current configuration can be defined as
\begin{equation}
\gmidShellCur_\alpha = \frac{\partial\posmidCur}{\partial\coordOrth_\alpha},
\quad \quad 
\gmidShellCur_3=\frac{\gmidShellCur_1\times\gmidShellCur_2}{\|\gmidShellCur_1\times\gmidShellCur_2\|}.
\label{eq: currentbases}
\end{equation}
These definitions allow us to express the positions, $\posIni$ and $\posCur$, of a point in the initial and the deformed configuration, respectively, as
\begin{align}
    \posIni &= \posmidIni + \coord_3 \gmidShellIni_3, \\
     \posCur &= \posmidCur + \coord_3 \lambda_3 \gmidShellCur_3,
\end{align}
where $\coord_3 (-\frac{\thickIni}{2} \leq \coord_3 \leq \frac{\thickIni}{2})$ denotes the perpendicular distance of the point to the midsurface with $\thickIni$ being the shell thickness in the reference configuration and $\lambda_3=h/H$ denotes the thickness stretch with $h$ and $H$ being the current and initial thickness. We note that the thickness stretch represents an additional unknown quantity, that is obtained by the use of incompressibility and plane stress conditions, as described later in Section~\ref{sec:Incompressibility}.

The basis vectors, $\gShellIni_\alpha=\frac{\partial \posIni}{\partial \coordOrth_\alpha}$, at a general point in the shell continuum are
\begin{align}
\gShellIni_\alpha &= \gmidShellIni_\alpha + \coord_3 \gmidShellIni_{3,\alpha}, \label{eq: basisShell}\\
\gShellIni_3 &= \gmidShellIni_3. \label{eq: normalShell}
\end{align}
The metrics are obtained as
\begin{align}
    \gShellInicomp_{\alpha \beta} &= \gmidShellInicomp_{\alpha \beta} +2 \coord_3 \curvInicomp_{\alpha \beta} +\coord_3^2 \gmidShellIni_{3,\alpha} \cdot \gmidShellIni_{3,\beta}, \label{eq: shellgabquad}\\
    \gShellInicomp_{\alpha 3} &= \gShellInicomp_{3 \alpha} =\gmidShellIni_\alpha \cdot \gmidShellIni_3 +\coord_3 \gmidShellIni_{3,\alpha} \cdot \gmidShellIni_3=0, \\
    \gShellInicomp_{33}&=\gmidShellInicomp_{33}=1,
\end{align}
where $\gmidShellInicomp_{\alpha \beta}$ are the components of the covariant metrics (first fundamental form) in the initial configuration given as
\begin{align}
\label{eq: covariantmetricIni}
    \gmidShellInicomp_{\alpha \beta} = \gmidShellIni_\alpha \cdot \gmidShellIni_\beta,
\end{align}
and $\curvInicomp_{\alpha \beta}$ are the components of the curvature (second fundamental form) of the midsurface in the initial configuration given as
\begin{align}
    \curvInicomp_{\alpha \beta} = \frac{1}{2} (\gmidShellIni_\alpha \cdot \gmidShellIni_{3,\beta} + \gmidShellIni_\beta \cdot \gmidShellIni_{3,\alpha}).
\end{align}
We linearize strain through the thickness and therefore neglect the quadratic term in equation~(\ref{eq: shellgabquad}) to obtain~\cite{kiendl2015isogeometric}:
\begin{align}
    \gShellInicomp_{\alpha \beta} &= \gmidShellInicomp_{\alpha \beta} +2 \coord_3 \curvInicomp_{\alpha \beta}. \label{eq: shellGab}
\end{align}
Similarly, using the basis vectors in the current configuration $\gShellCur_\alpha=\frac{\partial \posCur}{\partial \coordOrth_\alpha}$, the metrics in the current configuration are
\begin{align}
    \gShellCurcomp_{\alpha \beta} &= \gmidShellCurcomp_{\alpha \beta} +2 \coord_3 \curvCurcomp_{\alpha \beta}, \label{eq: shellgab}
\end{align}
with
\begin{align}
\label{eq: covariantmetric}
    \gmidShellCurcomp_{\alpha \beta} = \gmidShellCur_\alpha \cdot \gmidShellCur_\beta, \qquad \qquad     \curvCurcomp_{\alpha \beta} = \frac{1}{2} (\gmidShellCur_\alpha \cdot \gmidShellCur_{3,\beta} + \gmidShellCur_\beta \cdot \gmidShellCur_{3,\alpha}). 
\end{align}
This expression shows that $\gShellCurcomp_{\alpha \beta}$ are not components of the unit tensor in the initial configuration for curved surfaces (i.e. where $\curvInicomp_{\alpha \beta}\neq 0$). Therefore, following Flores and O\~{n}ate~\cite{flores2005improvements}, we approximate the covariant components 
\begin{align}
    \gShellCurcomp_{\alpha \beta} = \gmidShellCurcomp_{\alpha \beta} +2 \coord_3 \changeCurvcomp_{\alpha \beta}, \label{eq: shellgab_final}
\end{align}
where $\changeCurvcomp_{\alpha \beta}$ denotes the change in curvature of the midsurface and is 
\begin{align}
\label{eq: curvChange}
    \changeCurvcomp_{\alpha \beta}=\curvCurcomp_{\alpha \beta}-\curvInicomp_{\alpha \beta}. 
\end{align}
We remark that this approximation, while not strictly necessary, facilitates the use of standard constitutive material models (e.g., equation~(\ref{eq:SEF-MR})) by enforcing that the metrics in initial configuration are components of unit tensor, and therefore, $\firstInv=3$ in the initial configuration.

\noindent Using the basis vectors defined in equations~(\ref{eq: basisShell}) and~(\ref{eq: normalShell}) and their counterparts in the deformed configuration, $\gShellCur_i$, the deformation gradient is
\begin{align}
\FShell=\frac{\partial\posCur}{\partial\posIni}
=\frac{\partial\posCur}{\partial\coordOrth_\alpha}\otimes\frac{\partial\coordOrth_\alpha}{\partial\posIni}
=\gShellCur_i\otimes\gShellIni^i,\label{eq:def_gradient_shell}
\end{align}
and the right Cauchy-Green strain tensor is
\begin{align}
\CShell=\FShell^T \FShell=\gShellCurcomp_{ij} \gShellIni^i \otimes \gShellIni^j.\label{eq: CShell}
\end{align}
Equation~(\ref{eq: CShell}) defines the right Cauchy-Green strain tensor that will be used to obtain the elastic strain energy for the shell formulation in Section~\ref{sec:ConsModel}.

\subsubsection{Continuum formulation}\label{sec:continuum}
\begin{figure}[!htbp]
	\centering
	\includegraphics[width=0.8\textwidth]{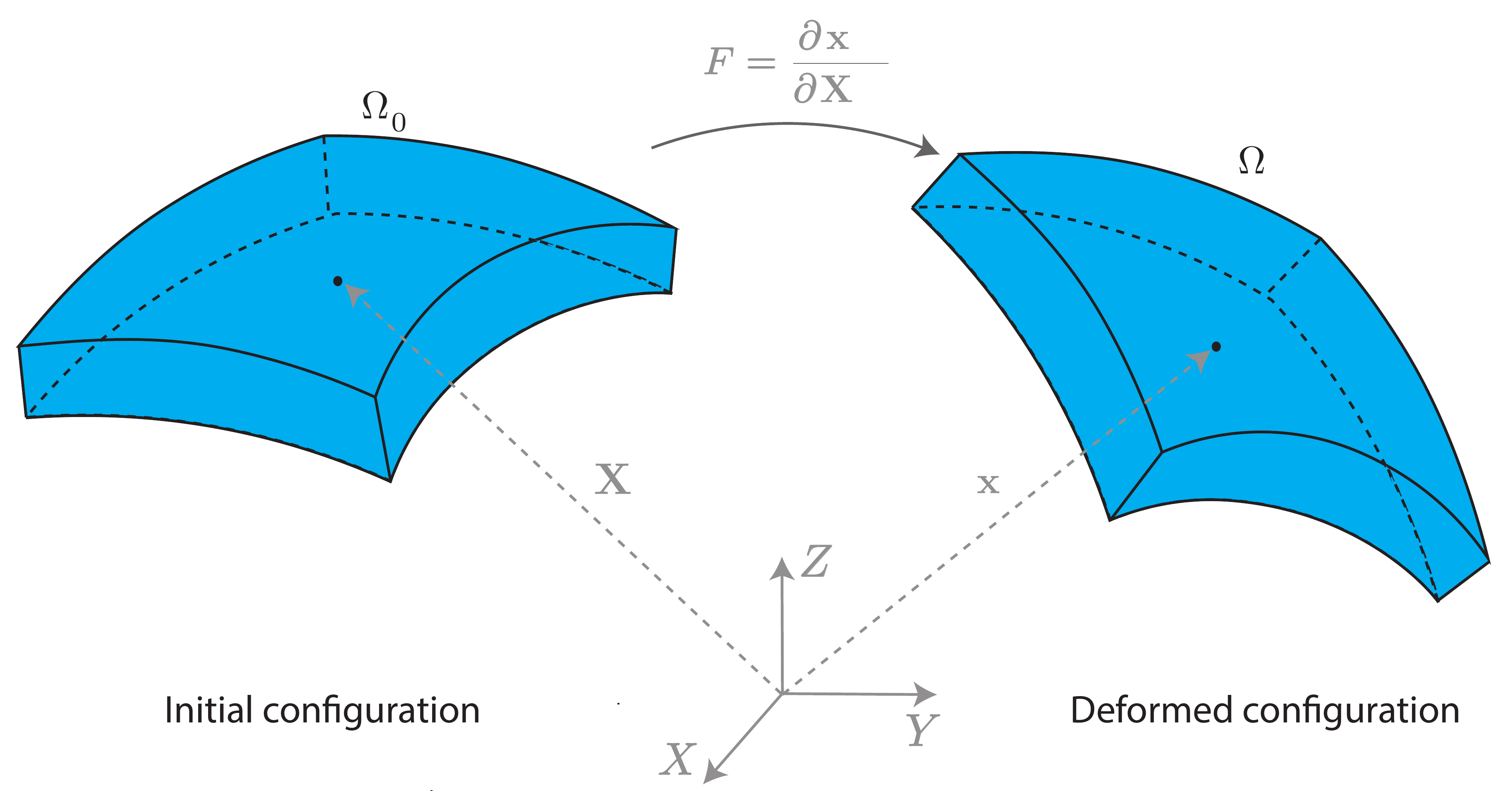}
	\caption{Schematic drawing illustrating the initial and deformed configuration for continuum formulation.}
	\label{fig:solid_schematic}
\end{figure}

Next, we describe the 3D nonlinear continuum formulation employed in this work. Referring to Figure~\ref{fig:solid_schematic},
let $\Omega_0\subset\mathbb{R}^3$, be an open, bounded and connected domain which represents the reference (or undeformed) configuration of the vessel wall tissue.
The deformation of the body $\Omega_0$ is defined through the mapping $\vect{\phi} :\Omega_0\rightarrow\mathbb{R}^3$ that
links a material point from the reference configuration $\vect{X}\in\Omega_0$ to a point in the deformed configuration $\vect{x}\in\Omega$ according to $\vect{x} = \vect{\phi}\left(\vect{X}\right)$ and $\Omega=\vect{\phi}(\Omega_0)$.  
Associated with the  mapping $\vect{\phi}$, the deformation gradient tensor $\vect{F}$  and the right Cauchy-Green strain tensor $\vect{C}$ are defined as
\begin{equation}\label{eqn:F1}
\FShell=\frac{\partial\posCur}{\partial\posIni}, \quad \quad \vect{C} =  \vect{F}^T\vect{F}.
\end{equation}
%
Equation~(\ref{eqn:F1}) provides the strain measure that allows the definition of constitutive models, as described later in Section~\ref{sec:ConsModel}. We remark that, in contrast to the membrane and shell formulations that enforce the incompressibility constraint, the continuum formulation enforces quasi-incompressibility (see Section~\ref{sec:Incompressibility} for further details).
We further note that, as opposed to the shell and membrane formulations, the 3D formulation requires no assumptions concerning through-thickness variation of kinematic quantities, other than those implicit in the choice of interpolation. 
The strategy for generation of solid mesh as well as the different interpolations considered for the 3D continuum formulation are discussed later in Section~\ref{sec: Discretization}.

\subsection{Constitutive Models}
\label{sec:ConsModel}
Next, we describe the two different constitutive models considered in this work. We note that the expressions for the elastic strain energy provided below are standard representations of these constitutive models; however, these expressions are typically modified by introducing Lagrange multipliers or penalty terms for the enforcement of incompressibility condition, as described later in Section~\ref{sec:Incompressibility}. 

\paragraph{\textbf{Mooney-Rivlin:}} The elastic strain energy for Mooney-Rivlin model is 
\begin{equation}
    \esef = \firstmat (\firstInv -3) + \secondmat (\secondInv - 3),\label{eq:SEF-MR}
\end{equation}
where $\firstmat$, $\secondmat$ are the material parameters and $\firstInv$, $\secondInv$ are the first and second principal invariants of right Cauchy-Green tensor given as
\begin{align}
    \firstInv = \textrm{tr}(\CShell), \qquad
    \secondInv =\frac{1}{2} \big[\textrm{tr}(\CShell)^2 -\textrm{tr}(\CShell^2)\big].
\end{align}
For the special case of $\secondmat=0$, this model reduces to the incompressible Neo-Hookean model.

\paragraph{\textbf{Four-fiber family model:}} The four-fiber family model characterizes arterial wall as a composite structure comprising an elastin-dominated isotropic matrix and four families of embedded collagen fibers. The elastic strain energy is~\cite{eberth2011evolving,roccabianca2014quantification,ferruzzi2011mechanical}
\begin{equation}
    \esef = \frac{\isomat}{2} (\firstInv -3) +\sum_{k=1}^4 \frac{\firstmat^k}{4 \secondmat^k} \Big[ e^{\secondmat^k [(\fiberstretch^k)^2-1]^2}-1 \Big],\label{eq:SEF-4fib}
\end{equation}
where $\isomat$, $\firstmat^k$, $\secondmat^k$ are material parameters, $\firstInv$ is again the first invariant of right Cauchy-Green tensor and $\fiberstretch^k=\sqrt{\fiberdirection^k \cdot \CShell \fiberdirection^k}$ is the stretch experienced by the $k$-th fiber family that is oriented along the direction $\fiberdirection^k = [0,\sin\fiberangle^k, \cos \fiberangle^k]$ in the reference configuration. The four-fiber families are considered to be aligned along $\fiberangle^1=0^{\circ}$ (axial family), $\fiberangle^2=90^{\circ}$ (circumferential family) and $\fiberangle^{3,4}=\pm \fiberangle$ (symmetric diagonal families), where $\fiberangle$ is a free parameter. The four-fiber family model has recently been recommended as the first choice constitutive model to simulate arterial deformations that lie outside the range of experimental tests~\cite{schroeder2018predictive} and has been extensively employed to describe the behavior of both human and murine arterial tissue~\cite{ferruzzi2011mechanical,roccabianca2014quantification}.

\subsubsection{Enforcement of incompressibility and plane stress}
\label{sec:Incompressibility}
In this section, we briefly discuss the strategies to enforce incompressibility constraint for each formulation considered in this work. Additionally, we discuss the plane stress condition employed in the membrane and shell formulations.

\paragraph{\textbf{Membrane formulation:}} Here, we follow the strategy suggested by Lu~\etal~\cite{lu2008inverse} where the membrane deformation is viewed as components of 3D motion. The corresponding 3D version of the in-plane right Cauchy-Green strain tensor, equation~(\ref{eq: CMemb}), is expressed as
\begin{align}
\label{eq: CMembFull}
\CShell=\gMembCurComp_{\alpha \beta} \gMembIni^\alpha \otimes \gMembIni^\beta + \lambda^2_3 \gMembIni_3 \otimes \gMembIni_3,
\end{align}
where $\lambda_3=h/H$ denotes the thickness stretch with $h$ and $H$ being the current and initial thickness of the membrane, and $\gMembIni_3$ denotes the unit normal vector to the membrane surface. Using equation~(\ref{eq: CMemb}) and (\ref{eq: CMembFull}), the 3D Jacobian determinant can be expressed as
\begin{equation}
\label{eq: jacobian}
    \JShell=\sqrt{\mathrm{det}(\CShell)}=\bar{J} \lambda_3,
\end{equation}
where $\bar{J}=\sqrt{\frac{\|\gmidShellCur_1\times\gmidShellCur_2\|}{\|\gmidShellIni_1\times\gmidShellIni_2\|}}$ denotes the in-plane Jacobian determinant. The incompressibility constraint is enforced by setting $\JShell=1$, which in combination with equation~(\ref{eq: jacobian}),
yields 
\begin{align}
\label{eq: thicknesstretch}
  \lambda_3=1/\bar{J}.    
\end{align}
This allows us to express the 3D strain invariants in terms of the in-plane invariants, as
\begin{align}
    \firstInv=\overline{I}_C+\overline{II}_C^{-1}, \qquad \secondInv=\overline{I}_C \overline{II}_C^{-1}+\overline{II}_C,
\end{align}
where
\begin{align}
     \overline{I}_C = \textrm{tr}(\CMemb), \qquad
    \overline{II}_C =\frac{1}{2} \big[\textrm{tr}(\CMemb)^2 -\textrm{tr}(\CMemb^2)\big].
\end{align}

Therefore, the 3D right Cauchy-Green strain tensor (equation~(\ref{eq: CMembFull})) and consequently, the elastic strain energy (equations~(\ref{eq:SEF-MR}) and ~(\ref{eq:SEF-4fib})) can be completely described in terms of in-plane quantities. Furthermore, it can be shown that, under plane stress condition, the in-plane components of the second Piola-Kirchoff stress are given as
    $S^{\alpha \beta} = 2\frac{\partial \esef(\CMemb)}{\partial \gMembCurComp_{\alpha \beta}}$.
We refer the reader to Lu~\etal~\cite{lu2008inverse} for further details on constitutive modeling in membrane formulation.

\paragraph{\textbf{Rotation-free shell formulation:}} Here, we follow the strategy in ~\cite{kiendl2009isogeometric, Nama2020} where incompressibility is enforced by augmenting the elastic strain energy function $\esef(\CShell)$ with a constraint term using a Lagrange multiplier $\lmShell$ as
\begin{equation}
    \tsef = -\lmShell (\JShell-1) + \esef(\CShell).\label{eq:SEF}
\end{equation}
Here, we again follow the same arguments as above to express the thickness stretch $\lambda_3=h/H$ in terms of in-plane quantities, as in equation~(\ref{eq: thicknesstretch}). Furthermore, the additional unknown $\lmShell$ introduced in equation~(\ref{eq:SEF}) can be determined using a plane stress condition. A detailed description of this strategy can be found in Kiendl~\etal~\cite{kiendl2015isogeometric} and our recent work~\cite{Nama2020}.

\paragraph{\textbf{3D Continuum formulation:}} Before discussing the enforcement of incompressibility, we note that the elastic strain energy $\esef$ for the Mooney-Rivlin model in equation \eqref{eq:SEF-MR} is not stress free in the origin, namely, $\boldsymbol{S}:=\left.2\frac{\partial\psi_{\text{el}}}{\partial{\boldsymbol{C}}}\right\vert_{\boldsymbol{C}=\boldsymbol{I}}\neq \boldsymbol{0}$. This issue can be addressed by conveniently modifying equation \eqref{eq:SEF-MR} as
\begin{equation}
    \esef = \firstmat (\firstInv -3) + \secondmat (\secondInv - 3) - 2\left(c_1+2c_2\right)\ln J.\label{eq:SEF-MR continuum}
\end{equation}

\noindent Similarly, for the four-fiber family model in equation \eqref{eq:SEF-4fib}, the stress-free condition in the origin yields
    \begin{equation}
    \esef = \frac{\isomat}{2} (\firstInv -3) +\sum_{k=1}^4 \frac{\firstmat^k}{4 \secondmat^k} \Big[ e^{\secondmat^k [(\fiberstretch^k)^2-1]^2}-1 \Big] - c\ln{J}.\label{eq:SEF-4fib coorected}
\end{equation}

With regards to incompressibility, there are various possibilities to enforce it. One approach is to enforce full incompressibility, as in equation \eqref{eq:SEF}, by introducing a Lagrange multiplier $p$ that weakly enforces $J=1$. However, for the 3D continuum formulation, this entails the use of a mixed formulation with Finite Element spaces for both displacements and the Lagrange multiplier $p$ which comply with the \emph{inf-sup} or LBB condition~\cite{boffi2013mixed}. Therefore, in this work, we have instead employed a displacement-based formulation where the incompressibility of the material is relaxed and is enforced in a penalty-type manner by adding a volumetric term $U(J)$~\cite{bonet1997nonlinear}
\begin{equation}
    \psi = \psi_{\text{el}} + U(J);\qquad  U(J)=\frac{\lambda}{2}(J-1)^2.
\end{equation}
In this work, we employ a sufficiently large value of $\lambda$ yielding a good approximation of the incompressibility of the material with $\lambda\approx 10^2\times 2\left(c_1+c_2\right)$ for the model in equation~\eqref{eq:SEF-MR continuum} and $\lambda=10^2\times c$ for the model in equation~\eqref{eq:SEF-4fib coorected}.
 
\subsection{Discretization}
\label{sec: Discretization}
\begin{figure}[!htbp]
	\centering
	\includegraphics[width=1.0\textwidth]{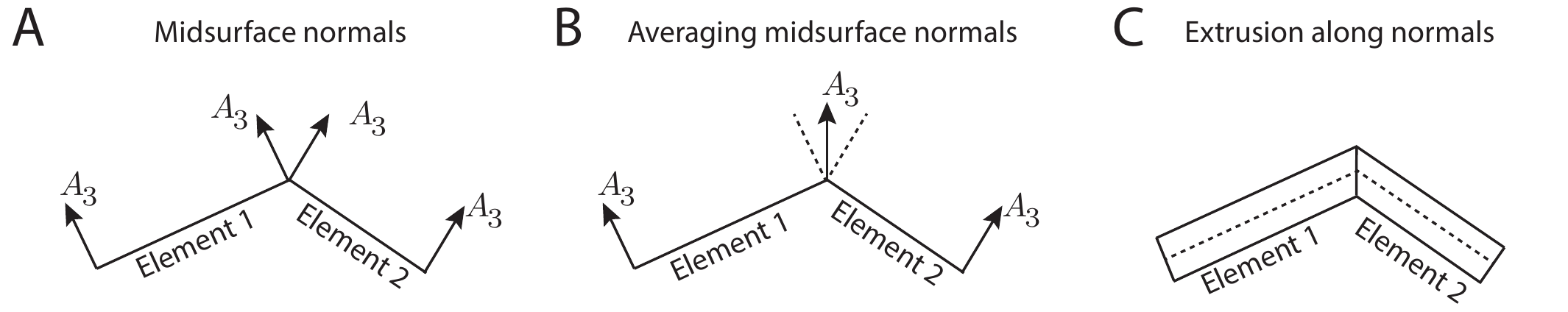}
	\caption{Generation of a solid mesh using extrusion of the midsurface mesh}
	\label{fig:extrusion}
\end{figure}

\begin{figure}[!htbp]
	\centering
	\includegraphics[width=0.6\textwidth]{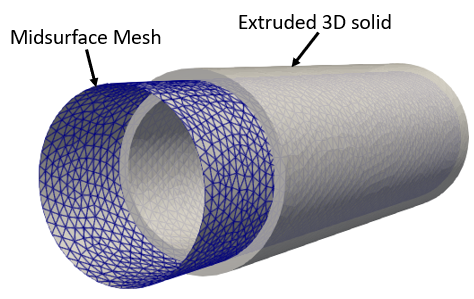}
	\caption{Extrusion of the midsurface mesh of a cylinder to generate a 3D solid cylindrical geometry. The resulting solid geometry has been clipped to reveal the initial midsurface mesh.}
	\label{fig:cylinder_extrusion}
\end{figure}

To compare the performance of different formulations under the closest possible conditions, 
we have chosen a particular type of finite element discretization for the continuum formulation. Specifically, the solid mesh consists of wedge (or triangular prismatic) elements~\cite{zienkiewicz2005finite} generated by performing an extrusion of the midsurface triangular elements used for the membrane and shell formulations.
This extrusion process comprises the following three steps: (1) Computation of the normal vector in the undeformed configuration, $\vect{A}_3$ at each node within every element, (2) Generation of continuous distribution of $\vect{A}_3$ across elements by local averaging, and (3) Extrusion of triangular elements along the continuous normal vectors $\vect{A}_3$, both inwards and outwards of the midsurface. This process is summarised in Figure~\ref{fig:extrusion}. 
Figure~\ref{fig:cylinder_extrusion}
shows the results of this extrusion strategy, applied on a midsurface mesh of a cylinder to generate a solid cylindrical geometry. Here, the resulting solid geometry has been clipped to reveal the initial midsurface mesh.

While both the P1 wedge and P2 wedge formulations employ the same solid mesh, the difference in interpolation strategies results in different number of degrees of freedom. Specifically, the P1 wedge formulation employs a linear interpolation of displacement, both in the midsurface plane and through-thickness. Consequently, for a wedge element with six nodes, it results in twice the number of degrees of freedom compared to the membrane or rotation-free shell formulation. In contrast, the P2 wedge formulation employs a quadratic interpolation of quantities both in the midsurface plane and through thickness, resulting in approximately 12 times degrees of freedom, compared to the membrane or rotation-free shell formulation. Table~\ref{tbl: Dofs} lists the resulting number of degrees of freedom for each geometry for different formulations.
\begin{table}[ht]
	\centering
\begin{tabular}{|p{70pt}|p{70pt}|p{70pt}|p{70pt}|p{70pt}|}
\hline
Geometry & Membrane  & Shell & P1 wedge & P2 wedge
\\ \hline
Cylinder & 8,601 & 8,601 & 17,202 & 102,420                                     \\  \hline
Bifurcation & \bf{--} & 48,945 & 97,890 & 586,134                               \\ \hline
Mouse-specific & \bf{--} & 63,843 & 127,686 & 764,865
\\ \hline
\end{tabular}
\caption{Degrees of freedom for each geometry for different formulations}
\label{tbl: Dofs}
\end{table}

\section{Results}
Next, we present results for the comparison of different formulations in three geometries: (a) a cylindrical geometry, (b) an idealized bifurcation model comprising  a main artery and an orthogonal side branch, and (c) a mouse-specific arterial model featuring multiple branches. These examples have been carefully chosen to introduce increasing degree of complexity, both in terms of the geometry as well as the material properties. Table~\ref{tbl: Examples} provides a summary of the degree of complexity considered for each example.
\begin{table}[ht]
	\centering
\begin{tabular}{|p{70pt}|p{70pt}|p{80pt}|p{80pt}|p{70pt}|}
\hline
Geometry & Material Model  & Regionally Varying Thickness & Regionally Varying Properties & External Tissue Support  \\
\hline
Cylinder & Four-fibers & No & No & No                                   \\ \hline
Bifurcation & Neo-Hookean & Yes & No & No                                       \\ \hline
Mouse-specific & Four-fibers & Yes & Yes & Yes
\\ \hline
\end{tabular}
\caption{Summary of different example cases}
\label{tbl: Examples}
\end{table}
Specifically, in terms of geometrical complexity, the first example represents the simplest idealized cylindrical model. The second example introduces a bifurcation in the geometry while still employing idealized cylindrical representations of individual vessels. Lastly, the third example considers subject-specific arterial model featuring multiple branches.
Similarly, in terms of material properties, the first example represents the simplest case with uniform material properties and wall thickness. The second example employs uniform material properties with regionally varying wall thickness. Lastly, the third example considers subject-specific, regionally-varying values for both material properties and wall thickness. In addition, this example also considers external tissue support conditions, as described later in Section~\ref{sec:Results-Mouse}.

\subsection{Cylindrical Geometry}
First, we compare the performance of different formulations against the well-known axisymmetric solution for a cylindrical tube. Details of this solution can be found elsewhere~\cite{Holzapfel2000a,humphrey2013cardiovascular}. This solution provides, in the most general case, the static mechanical response of a thick axisymmetric tube under combined bending, distension, extension, and torsion. 

We consider a cylindrical tube with midsurface radius $R=8.6~\textrm{mm}$ and axial length $L=60~\textrm{mm}$. To assess the performance of different formulations for increasing wall thickness to radius ratio, we consider three different thickness to radius ratio $H/R=0.05$, $0.25$, and $0.5$, while keeping the midsurface radius fixed. The arterial wall is assumed to be characterized by a four-fiber family model with the material parameters listed in Supplementary Table S1. For the purpose of this example, the tube is subjected to an external pressure of $80~\textrm{mmHg}$ while the ends of the tube are completely fixed. The comparison is performed in the mid-plane slice along the axial direction. The axisymmetric solution is taken to be the true solution and is compared against the solutions obtained from four different implementations: membrane, shell, P1 wedge, and P2 wedge.

\begin{figure}[!htbp]
	\centering
	\includegraphics[width=1.0\textwidth]{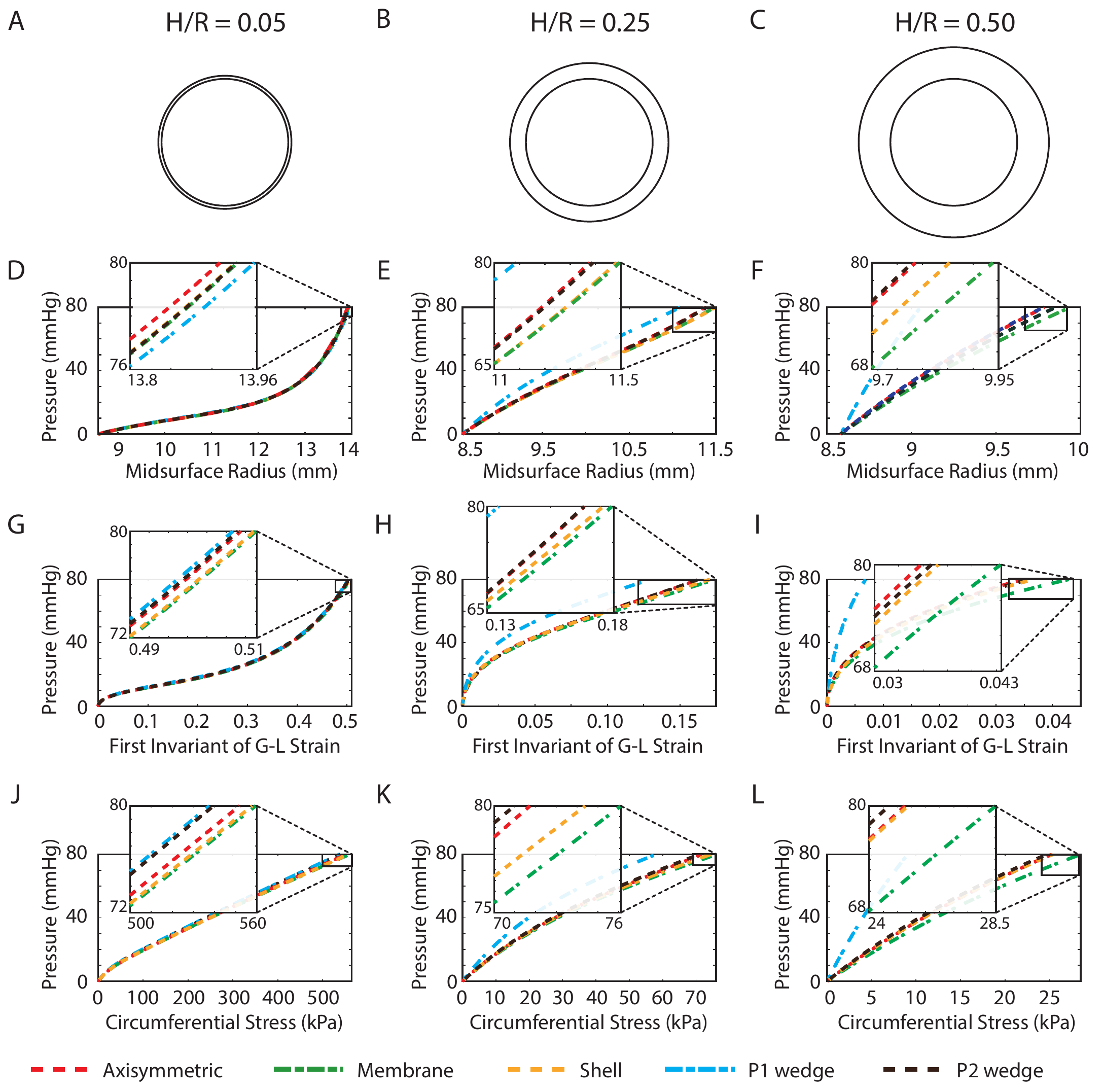}
	\caption{(A-C) Cross-sections of the cylindrical tube for three different thickness-to-radius ratios. (D-F), (G-H), and (J-L) compare the evolution of midsurface radius, first invariant of Green-Lagrange strain tensor, and circumferential Cauchy stress, respectively, with pressure for all formulations against the axisymmetric solution. 
	}
	\label{fig:cylinderinflation}
\end{figure}

Figure~\ref{fig:cylinderinflation}(A-C) 
show the cross section of the cylindrical tube for the three different thickness to radius ratios considered. Figure~\ref{fig:cylinderinflation}(D-L) compare the evolution of three different metrics with pressure: midsurface radius (Figure~\ref{fig:cylinderinflation}(D-F)), first invariant of Green-Lagrange strain tensor (Figure~\ref{fig:cylinderinflation}(G-I)), and circumferential Cauchy stress (Figure~\ref{fig:cylinderinflation}(J-L)).  

For $H/R=0.05$, all the four implementations yield excellent agreement with the axisymmetric solution. 2D formulations perform slightly better than the solid formulations with maximum errors of $1.38\%$, $1.08\%$, $2.59\%$, and $2.2\%$ for the membrane, shell, P1 wedge, and P2 wedge implementations, respectively. As the wall thickness is increased to $H/R=0.25$, it can be observed that the P1 wedge solution starts to deviate significantly from the axisymmetric solution. Specifically, the P1 wedge solution yields $10.67\%$ smaller displacement, $20.8\%$ lower value of $I_1$, and $19.2\%$ lower stress, compared to the axisymmetric solution. Membrane and shell solutions exhibit a smaller deviation from the axisymmetric  solution, with maximum errors of $6.4\%$ and $4.6\%$, respectively. P2 wedge solution yielded the best agreement with the axisymmetric solution, with maximum error of $1\%$. As the wall thickness is further increased to $H/R=0.5$, a similar trend is observed again: compared to the axisymmetric solution, the P1 wedge solution yields a significantly stiffer solution (maximum error=$80\%$), the membrane and shell exhibit smaller deviation with maximum errors of $22.8\%$ and $5.6\%$, respectively, while P2 wedge solution yielded the best agreement, with maximum error of $2.8\%$.

We note that all the numerical formulation employ an unstructured, non-axisymmetric mesh that might introduce differences from the axisymmetric solution. Further, as noted earlier, the differences in handling of incompressibility can introduce differences among 2D and 3D formulations. Nonetheless, these results indicate that the 2D formulations (membrane and shell) perform better than the linear continuum formulation (P1 wedge) for the same surface mesh as the wall thickness is increased. This is an interesting finding, given that the P1 wedge solution is obtained with twice the number of degrees of freedom compared to the 2D formulations. In contrast, P2 wedge formulation, using the highest number of degrees of freedom, yields the best agreement with the axisymmetric solution, except for the thinnest geometry ($H/R=0.05$). 

\subsection{Idealized Bifurcation Geometry}
\label{sec: bifurcation}
Next, we consider an idealized bifurcation model comprising a main artery and a side branch. The main vessel is a cylinder with radius $0.6~\mathrm{mm}$ while the side branch is a cylinder with radius $0.2~\mathrm{mm}$. The two branches meet at a right angle with a fillet radius $0.2~\mathrm{mm}$. The wall thickness is taken to be 10\% of the local radius, resulting in wall thickness of $0.06~\textrm{mm}$ and $0.02~\textrm{mm}$ for the main vessel and the branch, respectively, with a smooth variation of thickness in the blend region. Therefore, this example introduces two additional complexities compared to the cylindrical geometry: (a) a bifurcation region due to the presence of a side branch, and (b) regionally varying thickness due to different thickness of the main artery and the side branch. Without a loss of generality, we assume a Neo-Hookean material model for the vessel wall with $c_1=0.125~\textrm{MPa}$ and subject this geometry to a pressurization of $80~\mathrm{mmHg}$ with fixed ends boundary conditions. 
The membrane formulation failed to converge for this example, likely due to its inability to handle bending mode deformations in the blend region. Consequently, the comparison for this case is performed only among shell, P1 wedge, and P2 wedge formulations. 

\begin{figure}[!htbp]
	\centering
	\includegraphics[width=1.0\textwidth]{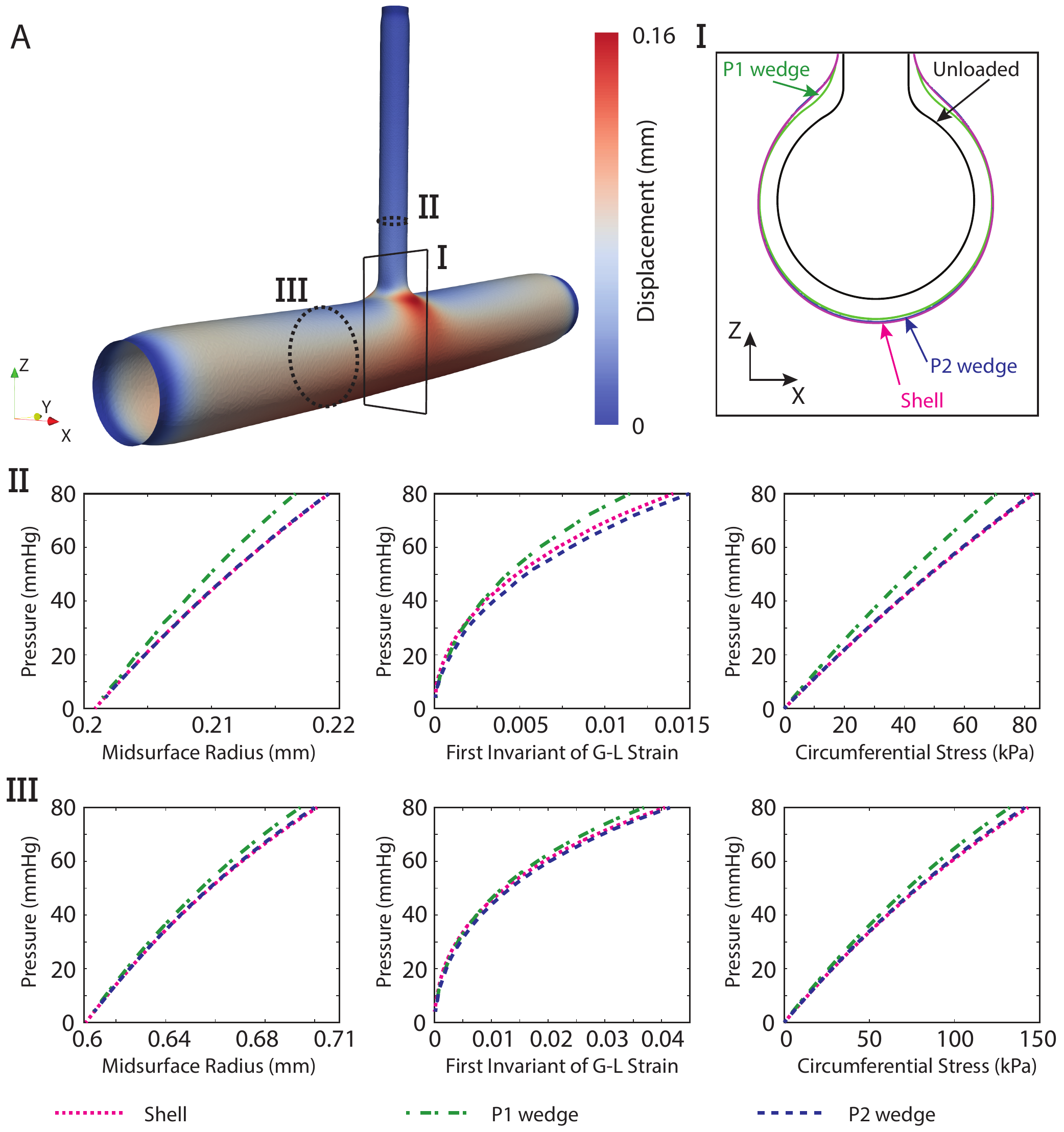}
	\caption{Comparison of shell, P1 wedge, and P2 wedge formulations for an idealized bifurcation geometry.}
	\label{fig:TBend}
\end{figure}

Figure~\ref{fig:TBend}(A)
shows the displacement in the final deformed configuration, obtained using the P2 wedge formulation, with maximum displacement observed near the bifurcation. Figure~\ref{fig:TBend}(I) compares the final deformed configuration at plane I, obtained using shell, P1 wedge, and P2 wedge formulations. Figure~\ref{fig:TBend}(I) also includes the corresponding unloaded configuration to indicate the extent of deformation during the pressurization.
It can be observed that the shell formulation agrees very well with the P2 wedge formulation. In contrast, the P1 wedge solution yields smaller displacements than the P2 wedge solution, particularly in the neck region.

Figure~\ref{fig:TBend}(II) and (III) compare the solution from different formulations at location II and III in Figure~\ref{fig:TBend}(A), respectively. We plot the evolution of three different metrics with pressure: midsurface radius, first invariant of Green-Lagrange strain tensor, and circumferential Cauchy stress. It can be observed that at both locations II and III, the shell solution is in excellent agreement with the P2 wedge solution (with maximum error of $0.7\%$) while the P1 wedge solution yields a stiffer solution, with maximum error of $14.5\%$.

This analysis revealed three significant findings: (1) The membrane formulation is unsuitable for general biomechanical analysis due to its inability to handle the bending mode deformation observed in geometries with branches, (2) the shell formulation yields excellent agreement with P2 wedge formulation, despite using only $8.4\%$ of the degrees of freedom (see Table~\ref{tbl: Dofs}), and (3) given the same midsurface mesh, the shell formulation captures the P2 wedge solution with superior accuracy than the P1 wedge solution, despite using $50\%$ fewer degrees of freedom.

\subsection{Mouse-specific Geometry}
\label{sec:Results-Mouse}
Next, we compare the performance of different formulations for a mouse-specific anatomy featuring multiple branches. Specifically, this geometrical model consists of a mouse aorta and four upper branches. The anatomy was obtained via non-invasive \invivo micro-CT and has been employed in our previous work~\cite{cuomo2019sex}. We model the aortic wall via a four-fiber family constitutive model with regionally varying material parameters and wall thickness distribution (see Table S1 in Supplementary text online). As illustrated in our prior work~\cite{Nama2020}, it is crucial to include the effect of surrounding tissue via external tissue support boundary conditions to avoid unphysiological deformations and buckling of the arterial wall. Therefore, following our prior work~\cite{Nama2020}, we model the effect of external tissue via an imposed traction on the vessel wall, $\CauchyStress \cdot \gmidShellCur_3 = -\stiffsupp \disp -\dampsupp \velocity -\thoracicpressure  \gmidShellCur_3$, where $\stiffsupp$ and $\dampsupp$ are the coefficients associated with simplified elastic and the viscoelastic responses of the external tissue, respectively, and $\thoracicpressure$ represents the extravascular pressure. For the current example, we use $\stiffsupp = 10^5~ \mathrm{g}/(\mathrm{mm}^2.\mathrm{s}^2)$, $\dampsupp = \thoracicpressure = 0$. Referring to Table~\ref{tbl: Examples}, this example introduces several additional complexities compared to previous examples by considering regional variation of material properties and wall thickness as well as the external tissue support conditions. Again, without a loss of generality, we subject this aortic geometry to pressurization of $80~\textrm{mmHg}$ with fixed ends boundary conditions.

\begin{figure}[!htbp]
	\centering
	\includegraphics[width=1.0\textwidth]{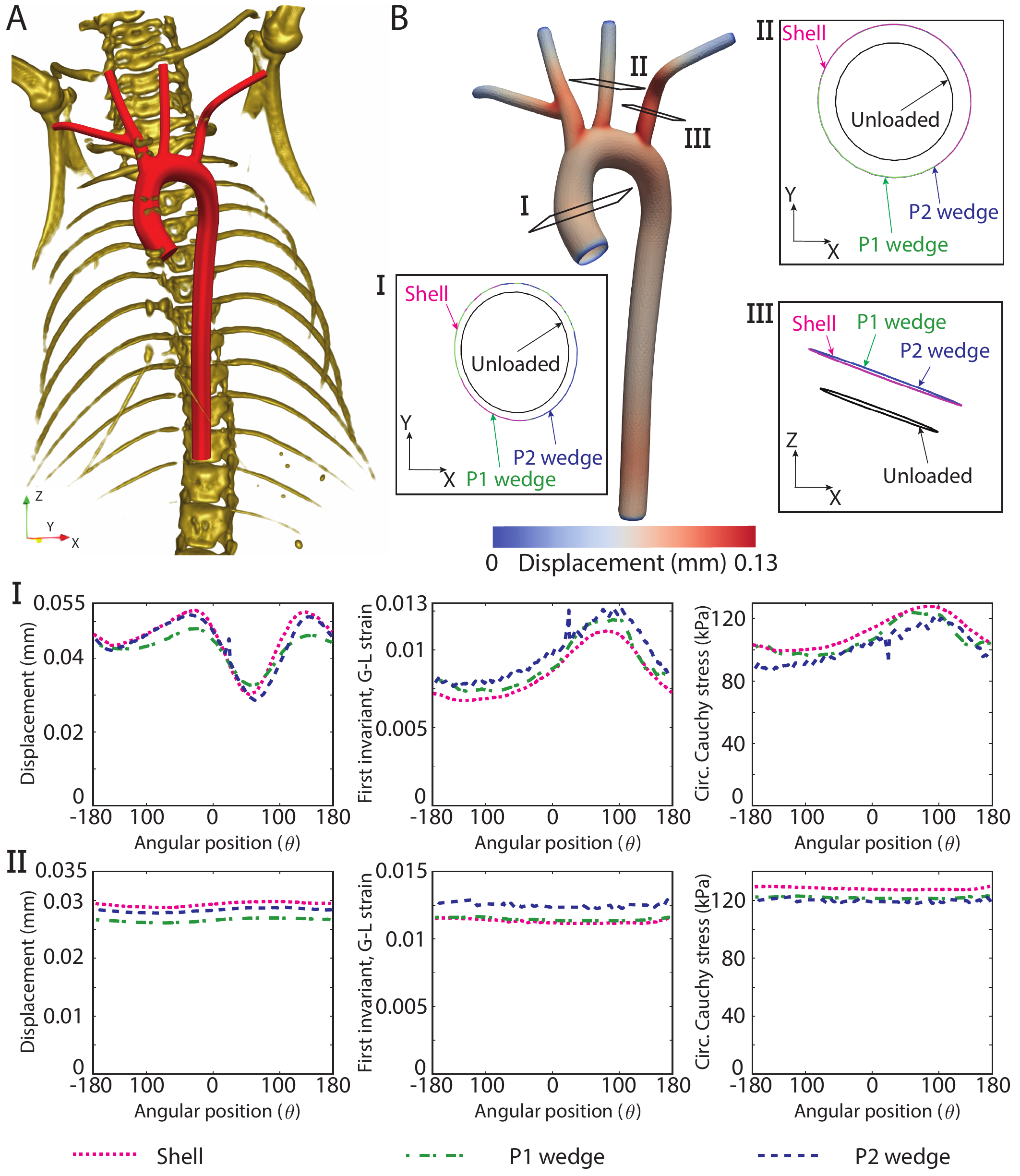}
	\caption{Comparison of shell, P1 wedge, and P2 wedge formulations for a mouse-specific geometry. (A) Volume rendered geometric model of the aortic anatomy of a 20 weeks old wild type female mouse. (B) plot of the final deformed configuration obtained via P2 wedge formulation, with enlarged versions comparing the deformed configurations predicted by three formulations at locations I, II, and III. Middle and the bottom row 
	plot the comparison of these formulations in terms of various kinematic metrics at the locations I and II indicated in panel (B).
    }
	\label{fig:Aorta}
\end{figure}

Figure~\ref{fig:Aorta}A
shows the volume rendered geometric model and the associated skeleton, obtained from the micro-CT data for a 20-week old wild-type mouse.
Figure~\ref{fig:Aorta}B shows the displacement in the final deformed configuration, obtained using the P2 wedge formulation. Panels I, II, and III in Figure~\ref{fig:Aorta}(B) compare the final deformed midsurface configuration obtained using different formulations at locations I, II, and III, respectively. The corresponding unloaded configuration of the midsurface has also been included to indicate the extent of deformation during the pressurization. The deformed configuration predicted by both the shell and P1 wedge formulation agrees excellently with the P2 wedge formulation at all three locations. It can also be observed from panel III of Figure~\ref{fig:Aorta}(B) that the deformation at location III is characterized by significant axial displacement. 

Next, we plot the comparison of these formulations in terms of various kinematic metrics at the locations I and II indicated in Figure~\ref{fig:Aorta}B. Specifically, we plot the circumferential variation of three different metrics: midsurface displacement, first invariant of Green-Lagrange strain tensor, and circumferential Cauchy stress, at locations I (mid row) and II (bottom row). Both the shell and P1 wedge formulations agree reasonably well with P2 wedge results, with maximum errors of $9.1\%$ and $5.1\%$, respectively. We note that this trend is different from the idealized bifurcation example considered in Section~\ref{sec: bifurcation}, where the shell formulation captured the P2 wedge solution with better accuracy than the P1 wedge formulation. This can be attributed to the fact that the current examples considers external tissue support that is applied differently in 2D and 3D formulations. Specifically, the external tissue support acts on the outer surface of the vessel wall in 3D formulations, while it acts on the wall midsurface in the shell formulation. From a  practical perspective, this implies that the choice of external tissue support parameters must be formulation-specific, to match the dynamic deformation data, if available. Nonetheless, despite this discrepancy, the results from all formulations agree reasonably well with each other. 

Overall, given that the shell formulation uses only $8.4\%$ and $50\%$ of degrees of freedom compared to the P1 wedge and P2 wedge formulation, respectively, (see Table~\ref{tbl: Dofs}) our results suggest that the shell formulation represents a computationally superior alternative, considering a balance between cost and accuracy, to the solid formulations. Furthermore, as noted in the Introduction, the shell formulation offers another advantage over 3D models by avoiding the need for creating an extruded mesh from the medical imaging data. 

\section{Discussion and Conclusions}
We presented a comparison of different nonlinear formulations for vascular biomechanics. Specifically, we considered three different formulations to represent the vessel wall: (i) a nonlinear membrane, (ii) a nonlinear, rotation-free shell, and (iii) a nonlinear 3D solid. For the 3D solid formulation, we considered two different implementations employing wedge elements with linear and quadratic interpolation, referred as P1 wedge and P2 wedge formulations, respectively. We presented comparison results in three geometries with increasing degree of complexity: (a) a cylindrical model, (b) an idealized bifurcation model comprising a main artery and a side branch, and (c) a mouse-specific arterial model featuring multiple branches. To ensure a fair comparison, the solid mesh was generated via an extrusion of the triangular mesh elements used for 2D formulations to obtain wedge (or triangular prismatic) elements.

For the cylindrical geometry, we compared the solution from the four implementations against the well-known axisymmetric solution, for three different thickness-to-radius ratios ($H/R=0.05, 0.25, 0.5$). For the thin wall case ($H/R=0.05$), all the formulations agreed excellently with the axisymmetric solution; 2D formulations performed slightly better than the solid formulations with maximum errors of $1.38\%$, $1.08\%$, $2.59\%$, and $2.2\%$ for the membrane, shell, P1 wedge, and P2 wedge implementations, respectively. However, as the wall thickness is increased relative to the radius ($H/R=0.25$), the P1 wedge solution exhibits significant deviation from the axisymmetric solution. The membrane and the shell formulations outperform the P1 wedge solution and yield results within $6.4\%$ and $4.6\%$, respectively, of the axisymmetric solution. The P2 wedge formulation provides the best agreement with the axisymmetric solution, with errors less than $1\%$. As the wall thickness is further increased relative to radius ($H/R=0.5$), a similar trend is observed, with the P1 wedge, membrane, shell, and P2 wedge formulations yielding errors of $80\%$, $22.8\%$, $5.6\%$, and $2.8\%$, respectively. It is interesting that the P1 wedge formulation consistently performs worse than the 2D formulations for all thickness to radius ratios, despite using twice the number of degrees of freedom.

Next, we compared results from different formulations in an idealized bifurcation model comprising of a main artery and a side branch, with regionally varying wall thickness. The membrane formulation failed to converge for this example, likely due its inability to handle bending mode deformations in geometries with branches. This observation indicates that a nonlinear membrane formulation is inappropriate for general vascular biomechanical analyses of complex geometries that routinely feature vessel bifurcations. Among the remaining formulations (shell, P1 wedge, and P2 wedge), the shell solution captured the P2 wedge solution with superior accuracy (maximum difference of $0.7\%$) than the P1 wedge solution (maximum difference of $14.5\%$). Again, these results point to the effectiveness of shell formulation for vascular biomechanics, given that it requires only $50\%$ and $8.4\%$ of the degrees of freedom for P1 wedge and P2 wedge formulations, respectively. 

Lastly, we compared results from the shell, P1 wedge, and P2 wedge formulations for a mouse-specific arterial model featuring multiple branches, while incorporating regional variation of material properties and wall thickness as well as external tissue support conditions. The results from the three formulations agree reasonably well with each other, with the shell and P1 wedge solutions yielding maximum differences of $9.1\%$ and $5.1\%$, respectively, relative to the P2 wedge solution. Given that the shell formulation employs significantly fewer degrees of freedom than the 3D formulations (see Table~\ref{tbl: Dofs}), these results again highlight the efficacy of shell formulation, considering a balance between cost and accuracy.

Overall, our results indicate that shell formulation represents an optimal compromise between the computational cost and accuracy for vascular biomechanics. Furthermore, given the lack of rotational degrees of freedom, the shell formulation is ideally-suited to serve as the structural model in fluid-structure interaction frameworks for vascular biomechanics. We are currently utilizing the shell formulation to develop a strongly coupled, monolithic, computationally efficient, nonlinear fluid-structure interaction framework via an adaption of the Coupled Momentum Method~\cite{figueroa2006coupled} to large displacements and nonlinear material models. Given the computational efficiency and accuracy of the shell formulation, this framework promises to provide a computationally efficient, yet accurate, alternative to the current fluid-structure interaction formulations that employ a 3D representation of the vessel wall~\cite{bathe1999fluid,hron2007fluid}.

\newpage
\bibliographystyle{nature}
\bibliography{bibliography}

\end{document}